\documentclass[conference]{IEEEtran}
\usepackage{setspace}
\usepackage{clrscode}
\usepackage{amsmath}
\usepackage{amssymb}
\usepackage[dvips]{graphicx}
\usepackage{epsfig}
\usepackage{hyperref}
\usepackage{bm}
\usepackage{subfigure}
\usepackage{verbatim}
\usepackage{titletoc}
\usepackage{extarrows}
\usepackage{fancyhdr}
\usepackage{geometry}
\usepackage{balance}

\newcommand{\figsize}{0.45}

\newtheorem{Lem1}{Proposition}

\newtheorem{Rem}{Remark}

\newtheorem{Def}{Definition}
\begin{document}

\title{Mimicking Full-Duplex Secure Communications for Buffer-Aided Multi-Relay Systems}
    \author{
\IEEEauthorblockN{Jiayu Zhou\IEEEauthorrefmark{1}, Deli Qiao\IEEEauthorrefmark{1}, and Haifeng Qian\IEEEauthorrefmark{2}}
\IEEEauthorblockA{\IEEEauthorrefmark{1}\small{School of Communication and Electronic Engineering, East China Normal University, Shanghai, China}}
\IEEEauthorblockA{\IEEEauthorrefmark{2}\small{School of Computer Science and Software Engineering, East China Normal University, Shanghai, China}}
\small{Email: 52191214003@stu.ecnu.edu.cn, dlqiao@ce.ecnu.edu.cn, hfqian@cs.ecnu.edu.cn}}

\maketitle

\begin{abstract}
This paper considers secure communication in buffer-aided cooperative wireless networks in the presence of one eavesdropper, which can intercept the data transmission from both the source and relay nodes. A new max-ratio relaying protocol is proposed, in which different relays are chosen for reception and transmission according to the ratio of the legitimate channels to the eavesdropper channels, so that the relay selected for reception and the relay selected for transmission can receive and transmit at the same time. It is worth noting that the relay employs a randomize-and-forward (RF) strategy such that the eavesdropper can only decode the signals received in the two hops independently. Theoretical analysis of the secrecy throughput of the proposed scheme is provided and the approximate closed-form expressions are derived, which are verified by simulations. Through numerical results, it is shown that the proposed scheme achieves a significant improvement in secrecy throughput compared with existing relay selection policies.
\end{abstract}

\section{Introduction}
Wireless communication technologies play an important role in military and civil applications, and their rapid developments have been promoting the evolution into the fifth generation (5G) communication \cite{5G1}, \cite{5G2}. However, the broadcast nature of wireless medium makes the communication over wireless networks susceptible to the interception attacks from unauthorized users (eavesdroppers), and thus guaranteeing the security of wireless communication is becoming an increasingly urgent demand \cite{demand}.

Traditionally, security issues are addressed by applying cryptographic methods, which utilize secret keys and encryption/decryption algorithms to provide secure data streams, in the upper layers of the network protocol stack \cite{encryption}. However, their applications may be limited by the inherent difficulty of secret key management and the increasingly powerful computation capability of the eavesdroppers. In the seminal work \cite{Wyner}, Wyner introduced the wiretap channel model and established the possibility of creating perfectly secure communication links without relying on secret keys. A rate at which information can be transmitted securely from the source to its intended destination is termed as achievable secrecy rate, and secrecy capacity is the maximal achievable secrecy rate. The secrecy capacity of the scalar Gaussian wiretap channel was analyzed in \cite{secrecy capacity}. Recently, physical layer (PHY) security techniques have been considered as a promising solution to guarantee everlasting secure communication for wireless networks by exploiting the inherent randomness of wireless channels and noise, and have received a lot of attention \cite{PHY}-\cite{outage}.

Of particular interest is the secure communication over relay channels, which is one of the fundamental building blocks of communication systems. In \cite{max-min-ratio}, the relay-eavesdropper channel was studied and different node cooperation strategies were analyzed. It has been shown that cooperative communication not only significantly improves the transmission capacity for wireless networks, but also provides an effective way to improve the secrecy capacity. On top of that, several relay strategies have been designed in literature \cite{AF/DF}-\cite{RF/DF}. Moreover, recent works have shown that use of buffers at the relays make it possible to store packets and transmit them in more favorable wireless conditions, which greatly improves the security performance of wireless communications \cite{Wan}- \cite{outage}. For instance, in \cite{Wan}, we have designed the link selection and power control policies for secure communications over a buffer-aided two hop communication link.

Note that a relay usually operates in either full-duplex (FD) or half-duplex (HD) mode. In FD relaying, the relays transmit and receive at the same time and frequency, at the cost of hardware complexity \cite{FD1}, \cite{FD2}. We consider a buffer-aided relay system with multiple HD relays in this paper. Inspired by the space full-duplex max-max relay selection (SFD-MMRS) scheme in \cite{mimick}, which mimics FD relaying with HD relays via link selection, we propose a new max-ratio relay selection scheme for secure transmission in HD randomize-and-forward (RF) buffer-aided relay networks in this paper. In the proposed scheme, we select different relays for reception and transmission according to the ratio of the legitimate channels to the eavesdropper channels. We analyze the secrecy throughput of the proposed scheme in independent and identically distributed (i.i.d.) Rayleigh fading channels and derive the approximate closed-form expressions. Numerical results in accordance with theoretical analysis show the superiority of the proposed scheme over the existing relay selection schemes.

The reminder of this paper is organized as follows. The system model and two existing max-ratio relay selection scheme for secure buffer-aided cooperative wireless networks are briefly introduced in Section II. In Section III, the relay selection policy is proposed, comprehensive analysis of the secrecy throughput is presented and the approximate closed-form expressions are derived. Numerical results are provided in Section IV. Finally, conclusions are drawn in Section V with some lengthy proofs in Appendix.

\section{Preliminaries}
\begin{figure}
    \centering
    \includegraphics[width=0.4\textwidth]{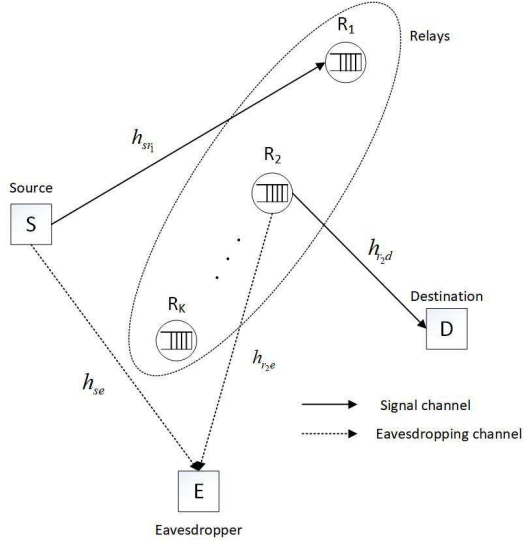}
    \caption{Illustration of system model.}
    \label{fig:figure1}
\end{figure}

\subsection{System Model}
We consider a two-hop wireless communication system consisting of one source node $S$, one destination node $D$, a set of $K$ relays $R_{1},...,R_{K}$ adopting the RF decoding strategy, and one eavesdropper $E$ which can intercept signals from both the source and relay nodes, as shown in Fig. \ref{fig:figure1}. Under the RF relaying strategy, the source and relay use different codebooks to transmit the secret message, so the eavesdropper cannot combine the data transmitted by source and relay \cite{RF/DF}, \cite{buffer}. We assume that each node is equipped with a single antenna and operates in the HD mode. We assume that there is a buffer of infinite length at each relay such that each relay can store the information received from the source and transmit it in later time.

We assume that there is no direct link between the source and destination due to high attenuation, and the communications can be established only via relays. The channel coefficients for $S-R_k$, $R_{k}-D$, $S-E$ and $R_{k}-E$ links at time $t$ are denoted as $h_{sr_{k}}(t)$, $h_{r_{k}d}(t)$, $h_{se}(t)$ and $h_{r_{k}e}(t)$, respectively. The channel is assumed to be stationary and ergodic. We consider the block fading, in which the channel coefficients remain constant during one time slot and vary independently from one to the other. In addition to fading, all wireless links are impaired by additive white Gaussian noise (AWGN) with variance $N_0$. Without loss of generality, we assume that the noise variances at the receiving nodes are equal to one, i.e., $N_{0}=1$. The source and relays are assumed to transmit with power $P_S$ and $P_R$ respectively.

Throughout this paper, we consider the case of i.i.d. Rayleigh fading for $S-R_i$, $R_{i}-D$, and $R_{i}-E$ links, which is a typical assumption to facilitate analysis \cite{mimick}. We assume that the mean of $S-R_k$, $S-E$, $R_{k}-D$ and $R_{k}-E$ channel gains are ${\mathbb E}[|h_{sr_{k}}|^{2}]=\gamma_{sr}$, ${\mathbb E}[|h_{se}|^{2}]=\gamma_{se}$, ${\mathbb E}[|h_{r_{k}d}|^{2}]=\gamma_{rd}$ and ${\mathbb E}[|h_{r_{k}e}|^{2}]=\gamma_{re}$, respectively, where ${\mathbb E}[\cdot]$ denotes the expectation.

\subsection{Existing Relaying Schemes}
In this part, we review two existing relay selection protocols for secure buffer-aided cooperative wireless networks.
\subsubsection{Max-Min-Ratio Relay Selection}
With the RF relaying strategy applied at the relays, the instantaneous secrecy rate for the buffer-aided multi-relay systems is obtained as \cite{max-min-ratio}
\begin{small}
\begin{align}
&C_{k}(t)=
\ \nonumber\\
&\Bigg[\frac{1}{2}{\rm log_{2}}\max\limits_{k\in\{1,...,K\}}\left\{{\rm min}\bigg(\frac{1+P_{S}|h_{sr_{k}}(t)|^{2}}{1+P_{S}|h_{se}(t)|^{2}},\frac{1+P_{R}|h_{r_{k}d}(t)|^{2}}{1+P_{R}|h_{r_{k}e}(t)|^{2}}\bigg)\right\}\Bigg]^{+},\label{max-min-ratio}
\end{align}
\end{small}
where $[\cdot]^{+}=\max\{\cdot,0\}$.
The secrecy throughput is given by ${\mathbb E}[C_{k}]$.
If the exact knowledge of the eavesdropping channels are available, the best relay node can be selected with the maximum $C_{k}(t)$. This scheme is termed as the max-min-ratio relay selection in this paper.
For convenience, the time index $t$ is ignored in the rest of the paper unless necessary.

\subsubsection{Max-Link-Ratio Relay Selection}
This protocol chooses the best link with the highest gain ratio among all available source-to-relay and relay-to-destination links \cite{max-link-ratio}.
If the exact knowledge of all channels, including the eavesdropping channels $h_{se}$ and $h_{r_{k}e}$, are available, the max-link-ratio selects the best relay $R_k$ as
\begin{small}
\begin{align}
k={\rm arg} \max\limits_{k\in\{1,...,K\}}\{\frac{\max\limits_{k\in\{1,...,K\}}\{|h_{sr_{k}}|^{2}\}}{|h_{se}|^{2}},\max\limits_{k\in\{1,...,K\}}\{\frac{|h_{r_{k}d}|^{2}}{|h_{r_{k}e}|^{2}}\}\}.\label{max-link-ratio}
\end{align}
\end{small}

\section{Full CSI At The Transmitters}
In this section, we assume the instantaneous channel state information (CSI) of legitimate channels (i.e., $|h_{sr_{k}}|^{2}$ and $|h_{r_{k}d}|^{2}$) are always known. Regarding the knowledge of eavesdropper CSI, we consider the perfect CSI case where the instantaneous eavesdropper CSI (i.e., $|h_{se}|^{2}$ and $|h_{r_{k}e}|^{2}$) are available in this paper.

Note however that, when the eavesdropper is passive and its behavior can not be monitored, the assumption of the exact CSI of the eavesdropper¡¯s link might be unrealistic. We also consider the partial CSI case where only the average gains of the eavesdropping channels are available. Since the derivations are similar, they are omitted in this paper due to the limit of space.

\subsection{Link Selection Policy}
Inspired by the SFD-MMRS scheme, if the channel gains of both the legitimate receiver and the eavesdropper are known at the transmitter, the best and the second best relay for reception $R_{r_{1}}$ and $R_{r_{2}}$ are selected respectively based on
\begin{small}
\begin{align}
&r_1={\rm arg} \max\limits_{k\in\{1,...,K\}}\left\{\frac{1+P_{S}|h_{sr_{k}}|^{2}}{1+P_{S}|h_{se}|^{2}}\right\},
\ \nonumber\\
&r_2=\arg \max_{\underset{k\neq r_1}{k\in \{1, \ldots, K\}}}  \left\{\frac{1+P_{S}|h_{sr_{k}}|^{2}}{1+P_{S}|h_{se}|^{2}}\right\},\label{reception}
\end{align}
\end{small}
and the best and the second best relay for transmission $R_{t_{1}}$ and $R_{t_{2}}$ are selected respectively according to

\begin{small}
\begin{align}
&t_1={\rm arg} \max\limits_{k\in\{1,...,K\}}\left\{\frac{1+P_{R}|h_{r_{k}d}|^{2}}{1+P_{R}|h_{r_{k}e}|^{2}}\right\},\notag
\end{align}
\end{small}
\begin{small}
\begin{align}
&t_2=\arg \max_{\underset{k\neq t_1}{k\in \{1, \ldots, K\}}}  \left\{\frac{1+P_{R}|h_{r_{k}d}|^{2}}{1+P_{R}|h_{r_{k}e}|^{2}}\right\}.\label{transmission}
\end{align}
\end{small}

Denote
\begin{small}
\begin{align}
&z_{r_1}=\frac{\max\limits_{k\in\{1,...,K\}}\left\{1+P_{S}|h_{sr_{k}}|^{2}\right\}}{1+P_{S}|h_{se}|^{2}},\ \nonumber\\
&z_{t_1}=\max\limits_{k\in\{1,...,K\}}\left\{\frac{1+P_{R}|h_{r_{k}d}|^{2}}{1+P_{R}|h_{r_{k}e}|^{2}}\right\},\ \nonumber\\
&z_{r_2}=\frac{\max\limits_{\underset{k\neq r_1}{k\in \{1, \ldots, K\}}}  \left\{1+P_{S}|h_{sr_{k}}|^{2}\right\}}{1+P_{S}|h_{se}|^{2}},\ \nonumber\\
&z_{t_2}=\max_{\underset{k\neq t_1}{k\in \{1, \ldots, K\}}}  \left\{\frac{1+P_{R}|h_{r_{k}d}|^{2}}{1+P_{R}|h_{r_{k}e}|^{2}}\right\}.\label{not symmetric}
\end{align}
\end{small}
Let
\begin{align}
Q=\min(z_{r_1}, z_{t_2})-\min(z_{r_2}, z_{t_1}).
\end{align}
Then, in the proposed imitating full-duplex max-max-ratio relay selection (IFD-MRRS) policy, the relays selected for reception $R_{\bar{r}_1}$ and transmission $R_{\bar{t}_1}$ are chosen as
\begin{align}
(R_{\bar{r}_1}, R_{\bar{t}_1})=
\begin{cases}
(R_{r_1}, R_{t_1}), \text{if} \quad r_1\neq t_1\\
(R_{r_1}, R_{t_2}), \text{if} \quad r_1= t_1 \; \text{and} \; Q>0\\
(R_{r_2}, R_{t_1}), \text{otherwise},\label{principle}
\end{cases}
\end{align}
which means if the best $S-R$ and the best $R-D$ channels do not share the same relay, i.e., $r_{1}\neq t_{1}$, we select the relay with the best $S-R$ channel for reception and the relay with the best $R-D$ channel for transmission. However, if the same relay has the best $S-R$ and $R-D$ channel, i.e., $r_{1}=t_{1}$, we choose the best bottleneck link, i.e., if $Q>0$, and select the relay with the best $S-R$ channel for reception and the relay with the second best $R-D$ channel for transmission. Conversely, if $Q<0$, we select the relay with the second best $S-R$ channel for reception and the relay with the best $R-D$ channel for transmission.

Note that the use of buffers make virtual full duplex operations at the relays possible when two different relays are selected for the reception and transmission (see, e.g., \cite{max-link-ratio} and \cite{mimick}).
%To be specific, the relay selected for reception stores the packet received from the source in its buffer, the packet stays in the buffer until the relay has been selected for transmission a sufficient number of times such that all older packets in the queue of the relay have been transmitted. And the relay selected for transmission transmits the first packet available in the queue of its buffer, which was received in a previous time slot, to the destination.

\subsection{Secrecy Throughput Analysis}
Considering that we employ the RF relaying strategy \cite{RF}, the eavesdropper cannot combine the data transmitted by source and relay at each time slot. Therefore, when either the source wishes to transmit confidential information to the relay or the relay sends private message to the destination, we can view as a single hop transmission in the presence of the interference from the other hop. Assuming that the eavesdropper employs decoding with successful interference cancellation and the decoding order at the eavesdropper is not known at the source or the relays, the maximum eavesdropping data rate is assumed to be upperbounded by $\log_{2}(1+P_{S}|h_{se}|^{2})$ for the link $S-E$, and $\log_{2}(1+P_{R}|h_{r_{k}e}|^{2})$ for the link $R_{k}-E$. If the relay $R_k$ is selected for the data transmission, the instantaneous secrecy rate of the first and second hop are lowerbounded by
\begin{small}
\begin{align}
C_{SR}\geq\left[{\rm log_{2}}\left(\frac{1+P_{S}|h_{sr_{k}}|^{2}}{1+P_{S}|h_{se}|^{2}}\right)\right]^{+},
\notag
\ \nonumber\\
C_{RD}\geq\left[{\rm log_{2}}\left(\frac{1+P_{R}|h_{r_{k}d}|^{2}}{1+P_{R}|h_{r_{k}e}|^{2}}\right)\right]^{+},
\end{align}
\end{small}
respectively. In the following, we adopt the lowerbound for the analysis, which represents the worst case and specifies the minimum throughput achievable.

Then the secrecy throughput for the buffer-aided multi-relay system is given by \cite{average throughput}
\begin{align}
C_s={\rm min}\{{\mathbb E}[C_{SR}], {\mathbb E}[C_{RD}]\},\label{secrecy throughput}
\end{align}
where ${\mathbb E}[C_{SR}]$ and ${\mathbb E}[C_{RD}]$ denote the average secrecy throughput of the $S-R$ and $R-D$ links, respectively.

The average secrecy rate of the first and second hop, $\overline{C}_{k}$, $k\in\{SR,RD\}$, can be expressed as \cite[(9)]{mimick}
\begin{align}
\overline{C}_{k}={\mathbb E}[C_{k}]=(1-p_{s})\overline{C}_{k,1}+p_{s}(\overline{C}_{k,21}+\overline{C}_{k,22}),
\end{align}
where $p_s$ denote the probability that $r_1=t_1$, $\overline{C}_{k,1}$ is the average throughput of the best channel of the first and second hop, $k\in\{SR,RD\}$, $\overline{C}_{k,21}$ and $\overline{C}_{k,22}$ denote the average throughput of the best channel and the second best channel of the first and second hop when $Q>0$ and $Q<0$, respectively.

Note that our selection policy involves eight independent channel coefficients, and hence finding the closed-form expressions is very tricky, if not intractable. So we derive the approximate closed-form expressions for the average secrecy throughput of the proposed scheme.

\begin{Lem1}
The average secrecy rate of the first hop ${\mathbb E}[C_{SR}]$ can be approximately expressed as
\begin{align}
&{\mathbb E}[C_{SR}]
\ \nonumber\\
&\approx(1-p_{s}){\mathbb E}[C_{SR,1}]+p_{s}(p_{12}{\mathbb E}[C_{SR,1}]+p_{21}{\mathbb E}[C_{SR,2}]),\label{ECSR}
\end{align}
where $p_s$ denotes the probability that $r_1=r_2$, $p_{12}$ denotes the probability that $Q>0$, $p_{21}$ denotes the probability that $Q<0$, ${\mathbb E}[C_{SR,1}]$ and ${\mathbb E}[C_{SR,2}]$ denote the average secrecy capacity of the best and the second best $S-R$ channel, respectively.
Similarly, the average secrecy rate of the second hop ${\mathbb E}[C_{RD}]$ can be approximately expressed as
\begin{align}
&{\mathbb E}[C_{RD}]
\ \nonumber\\
&\approx(1-p_{s}){\mathbb E}[C_{RD,1}]+p_{s}(p_{12}{\mathbb E}[C_{RD,2}]+p_{21}{\mathbb E}[C_{RD,1}]),\label{ECRD}
\end{align}
where ${\mathbb E}[C_{RD,1}]$ and ${\mathbb E}[C_{RD,2}]$ denote the average secrecy capacity of the best and the second best $R-D$ channel, respectively.
\end{Lem1}

\emph{Proof:} Based on (\ref{principle}), we can divide the time index $t$ into three cases correspondingly. If $r_{1}\neq t_{1}$, we select the relay with the best $S-R$ channel for reception and the relay with the best $R-D$ channel for transmission. We denote such indices as $t\in\Omega_{1}$. If $r_{1}=t_{1}$ , we need to determine whether Q is positive or negative, if $Q>0$, we select the relay with the best $S-R$ channel for reception and the relay with the second best $R-D$ channel for transmission. We denote such time indices as $t\in\Omega_{2}$. Inversely, if $Q<0$, we select the relay with the second best $S-R$ channel for reception and the relay with the best $R-D$ channel for transmission, and denote such time indices as $t\in\Omega_{3}$.
Let $N_{1}$ denote the number of times in N transmissions that $r_{1}=t_{1}$, and hence $N_{12}$ denote the number of time instances in $N_{1}$ transmissions that $Q>0$, and $N_{21}$ denote the number of time instances in $N_{1}$ transmissions that $Q>0$, i.e., $p_{s}=\frac{N_{1}}{N}$, $p_{12}=\frac{N_{12}}{N_{1}}$, and $p_{21}=\frac{N_{21}}{N_{1}}$.

Consider the first hop. The average secrecy rate is given by
\begin{small}
\begin{align}
&E[C_{SR}]
\ \nonumber\\
&=\lim\limits_{N\to \infty}\frac{1}{N}\sum\limits_{t=1}^{N}C_{SR}(t)
\ \nonumber\\
&=\lim\limits_{N\to \infty}\frac{1}{N}\left(\sum\limits_{t\in\Omega_{1}}C_{SR_{r_1}}(t)+\sum\limits_{t\in\Omega_{2}}C_{SR_{r_1}}(t)+\sum\limits_{t\in\Omega_{3}}C_{SR_{r_2}}(t)  \right)
\ \nonumber\\
&=\lim\limits_{N\to \infty}\frac{N-N_{1}}{N}\frac{1}{N-N_{1}}\sum\limits_{t\in\Omega_{1}}C_{SR}(t)
\ \nonumber\\
&+\frac{N_{1}}{N}\cdot\frac{N_{12}}{N_{1}}\frac{1}{N_{12}}\sum\limits_{t\in\Omega_{2}}C_{SR}(t)+\frac{N_{1}}{N}\cdot\frac{N_{21}}{N_{1}}\frac{1}{N_{21}}\sum\limits_{t\in\Omega_{3}}C_{SR}(t)
\ \nonumber\\
&=(1-p_{s}){\mathbb E}[C_{SR,1}]+p_{s}\cdot p_{12}{\mathbb E}[C_{SR,1}]+p_{s}\cdot p_{21}{\mathbb E}[C_{SR,2}].
\end{align}
\end{small}
The average secrecy rate of the second hop can be proved by the same logic. \hfill$\square$

\begin{Lem1}
Assume that full CSI of the eavesdropper's channels are known. Given $P_{S}$ and $P_{R}$, the detailed expressions for the terms in (\ref{ECSR}) and (\ref{ECRD}) can be expressed as follows:
\begin{small}
\begin{align}
&p_{s}=\frac{1}{K},\ \\
&{\mathbb E}[C_{SR,1}]=\sum\limits_{r=1}^{K}\binom K r (-1)^{r}\frac{e^{\frac{r}{P_{S}\gamma_{sr}}}}{\ln2}\bigg[-E_{1}\left(\frac{r}{P_{S}\gamma_{sr}}\right)
\ \nonumber\\
&+e^{\frac{1}{P_{S}\gamma_{se}}}E_{1}\left(\frac{r}{P_{S}\gamma_{sr}}+\frac{1}{P_{S}\gamma_{se}}\right)\bigg],\ \\
&{\mathbb E}[C_{SR,2}]=\sum\limits_{r=1}^{K-1}\binom {K-1}{r}(-1)^{r}\frac{e^{\frac{r}{P_{S}\gamma_{sr}}}}{\ln2}
\notag
\end{align}
\end{small}
\begin{small}
\begin{align}
&\left[-E_{1}\left(\frac{r}{P_{S}\gamma_{sr}}\right)+e^{\frac{1}{P_{S}\gamma_{se}}}E_{1}\left(\frac{r}{P_{S}\gamma_{sr}}+\frac{1}{P_{S}\gamma_{se}}\right)\right]
\ \nonumber\\
&+\binom {K-1}{K-2}\sum\limits_{r=0}^{K-1}\binom {K-1}{r}(-1)^{r}\frac{e^{\frac{r+1}{P_{S}\gamma_{sr}}}}{\ln2}
\ \nonumber\\
&\left[-E_{1}\left(\frac{r+1}{P_{S}\gamma_{sr}}\right)+e^{\frac{1}{P_{S}\gamma_{se}}}E_{1}\left(\frac{r+1}{P_{S}\gamma_{sr}}+\frac{1}{P_{S}\gamma_{se}}\right)\right],
\ \\
&p_{12}=\int_{0}^{\infty}f_{z_{r_2}}(z)(1-F_{z_{t_2}}(z))dz,
\end{align}
\end{small}
where
\begin{small}
\begin{align}
&f_{Z_{r_2}}(z)=\sum\limits_{r=0}^{K-1}\binom{K-1}{r}(-1)^{r}e^{-\frac{(z-1)r}{P_{S}\gamma_{sr}}}
\ \nonumber\\
&\left(\frac{-\frac{r}{P_{S}\gamma_{sr}}}{1+\frac{\gamma_{se}r}{\gamma_{sr}}z}-\frac{\frac{\gamma_{se}r}{\gamma_{sr}}}{\left(1+\frac{\gamma_{se}r}{\gamma_{sr}}z\right)^{2}}\right)
+\binom{K-1}{K-2}\sum\limits_{r=0}^{K-1}\binom{K-1}{r}
\ \nonumber\\
&(-1)^{r}e^{-\frac{(z-1)(r+1)}{P_{S}\gamma_{sr}}}\left(\frac{-\frac{r+1}{P_{S}\gamma_{sr}}}{1+\frac{\gamma_{se}(r+1)}{\gamma_{sr}}z}-\frac{\frac{\gamma_{se}(r+1)}{\gamma_{sr}}}{\left(1+\frac{\gamma_{se}(r+1)}{\gamma_{sr}}z\right)^{2}}\right),
\ \nonumber\\
&F_{Z_{t_2}}(z)=\left(1-\frac{e^{-\frac{z-1}{P_{R}\gamma_{rd}}}}{1+\frac{\gamma_{re}z}{\gamma_{rd}}}\right)^{K-1}\left(1+\binom {K-1}{K-2} \frac{e^{-\frac{z-1}{P_{R}\gamma_{rd}}}}{1+\frac{\gamma_{re}z}{\gamma_{rd}}}\right).\notag
\end{align}
\end{small}
\end{Lem1}
%\begin{small}
%\begin{align}
%&p_{12}= \int_{0}^{\infty}\Bigg[\sum\limits_{r=0}^{K-1}\binom{K-1}{r}(-1)^{r}e^{-\frac{(z-1)r}{P_{S}\gamma_{sr}}}
%\ \nonumber\\
%&\left(\frac{-\frac{r}{P_{S}\gamma_{sr}}}{1+\frac{\gamma_{se}r}{\gamma_{sr}}z}-\frac{\frac{\gamma_{se}r}{\gamma_{sr}}}{\left(1+\frac{\gamma_{se}r}{\gamma_{sr}}z\right)^{2}}\right)
%+\binom{K-1}{K-2}\sum\limits_{r=0}^{K-1}\binom{K-1}{r}
%\ \nonumber\\
%&(-1)^{r}e^{-\frac{(z-1)(r+1)}{P_{S}\gamma_{sr}}}\left(\frac{-\frac{r+1}{P_{S}\gamma_{sr}}}{1+\frac{\gamma_{se}(r+1)}{\gamma_{sr}}z}-\frac{\frac{\gamma_{se}(r+1)}{\gamma_{sr}}}{\left(1+\frac{\gamma_{se}(r+1)}{\gamma_{sr}}z\right)^{2}}\right)\Bigg]
%\ \nonumber\\
%&\left[1-\left(1-\frac{e^{-\frac{z-1}{P_{R}\gamma_{rd}}}}{1+\frac{\gamma_{re}z}{\gamma_{rd}}}\right)^{K-1}\left(1+\binom {K-1}{K-2}\frac{e^{-\frac{z-1}{P_{R}\gamma_{rd}}}}{1+\frac{\gamma_{re}z}{\gamma_{rd}}}\right)\right]dz.
%\end{align}
%\end{small}

\begin{Rem}
Note that ${\mathbb E}[C_{SR}]$ can be obtained by substituting the above equations into (\ref{ECSR}).

Similarly, we can obtained the expressions for ${\mathbb E}[C_{RD}]$.
\end{Rem}

\emph{Proof:} The probability that the best $S-R$ and the best $R-D$ channels share the same relay is $p_{s}=\frac{1}{K}$ \cite{mimick}, it follows directly from the fact that the channels for both $S-R$ and $R-D$ links are i.i.d. And it is obvious that we have $p_{21}=1-p_{12}$.
So, to compute the secrecy throughput of IFD-MRRS, we need to find ${\mathbb E}[C_{SR,1}]$, ${\mathbb E}[C_{RD,1}]$, ${\mathbb E}[C_{SR,2}]$, ${\mathbb E}[C_{RD,2}]$ and $p_{12}$.

\quad $Computation$ $of$ ${\mathbb E}[C_{SR,1}]$: In this case, we denote $z_{r_1}=\frac{1+P_{S}x}{1+P_{S}y}$, where $x=\max\{x_{k}\}$ with $x_{k}=|h_{sr_{k}}|^{2}, k\in\{1,2,...,K\}$, and $y=|h_{se}|^{2}$. To derive ${\mathbb E}[C_{SR,1}]$, we first compute the cumulative distribution function (CDF) of $z_{r_1}$. ${\mathbb E}[C_{SR,1}]$ can be obtained as
\begin{small}
\begin{align}
{\mathbb E}[C_{SR,1}]=&\sum\limits_{r=1}^{K}\binom K r (-1)^{r}\frac{e^{\frac{r}{P_{S}\gamma_{sr}}}}{\ln2}\bigg[-E_{1}\left(\frac{r}{P_{S}\gamma_{sr}}\right)
\ \nonumber\\
&+e^{\frac{1}{P_{S}\gamma_{se}}}E_{1}\left(\frac{r}{P_{S}\gamma_{sr}}+\frac{1}{P_{S}\gamma_{se}}\right)\bigg],\label{prop:ECSR1}
\end{align}
\end{small}
where $E_{1}(x)=\int_{x}^\infty(e^{-t}/t)dt, x>0$ is the exponential integral function.
See Appendix \ref{app:ECSR1} for the derivation of (\ref{prop:ECSR1}).

\quad $Computation$ $of$ ${\mathbb E}[C_{RD,1}]$: In this case, we denote $z_{t_1}=\max\limits_{k\in\{1,2,...,K\}}\{z_{k}\}$, where $z_{k}=\frac{1+P_{R}x_{k}}{1+P_{R}y_{k}}$ with $x_{k}=|h_{r_{k}d}|^{2}$ and $y_{k}=|h_{r_{k}e}|^{2}$. Since $x_{k}$ is independent of $y_{k}$, we have $f_{X_{k}Y_{k}}(x_{k},y_{k})=f_{X_{k}}(x_{k})f_{Y_{k}}(y_{k})$. To derive ${\mathbb E}[C_{SR,1}]$, we first compute the CDF of $z_{k}$,
then we can calculate the CDF of $z_{t_1}$. ${\mathbb E}[C_{RD,1}]$ is given by
\begin{small}
\begin{align}
&{\mathbb E}[C_{RD,1}]
\ \nonumber\\
&=\sum\limits_{r=1}^{K}\binom K r (-1)^{r}\frac{\gamma_{re}e^{\frac{r}{P_{R}}\left(\frac{1}{\gamma_{rd}}+\frac{1}{\gamma_{re}}\right)}}{\ln2\gamma_{rd}}
\ \nonumber\\
&\Bigg[-e^{-\frac{r}{P_{R}\gamma_{re}}}E_{1}\left(\frac{r}{P_{R}\gamma_{rd}}\right)
\ \nonumber\\
&+\sum\limits_{i=1}^{r}\bigg(\frac{(-1)^{i+1}\left(\frac{r}{P_{R}\gamma_{re}}\right)^{i+1}E_{1}\left(\frac{r}{P_{R}}\left(\frac{1}{\gamma_{rd}}+\frac{1}{\gamma_{re}}\right)\right)}{(i-1)!}
\notag
\end{align}
\end{small}
\begin{small}
\begin{align}
&+\frac{e^{-\frac{r}{P_{R}}\left(\frac{1}{\gamma_{rd}}+\frac{1}{\gamma_{re}}\right)}}{\left(\frac{\gamma_{re}}{\gamma_{rd}}+1\right)^{i-1}}
\sum\limits_{j=0}^{i-2}\frac{(-1)^{j}\left(\frac{r}{P_{R}\gamma_{re}}\right)^{j}\left(\frac{\gamma_{re}}{\gamma_{rd}}+1\right)^{j}}{(i-1)(i-2)...(i-1-j)}\bigg)\Bigg].\label{prop:ECRD1}
\end{align}
\end{small}
See Appendix \ref{app:ECRD1} for the derivation of (\ref{prop:ECRD1}).

\quad $Computation$ $of$ ${\mathbb E}[C_{SR,2}]$ and ${\mathbb E}[C_{RD,2}]$:
To compute ${\mathbb E}[C_{SR,2}]$ and ${\mathbb E}[C_{RD,2}]$, we need to compute the CDF of $z_{r_2}$ and $z_{t_2}$, so we consider the following theorem of order statistics:

\emph{Theorem 1 (\cite{order statistics})}: Let $Z_{1},...,Z_{n}$ be $n$ independent variates, each with cdf $F(z)$. Let $Z_{1},...,Z_{n}$ denote the increasing order of $Z_{1},...,Z_{n}$, i.e., $Z_{(1)}\leq Z_{(2)}\leq Z_{(n)}$. Let $F_{(r)}(z),(r=1,...,n)$ denote the cdf of the $r$ th order statistic $Z_{(r)}$. Then the cdf of $Z_{(r)}$ is given by
\begin{align}
F_{(r)}(z)=F^{r}(z)\sum\limits_{j=0}^{n-r}\binom {r+j+1}{r-1} [1-F(z)]^{j}. \label{general result}
\end{align}
Based on (\ref{general result}), we obtain
\begin{small}
\begin{align}
F_{Z_{r_2}}(z)=&\sum\limits_{r=0}^{K-1}\binom {K-1}{r} (-1)^{r}\frac{e^{-\frac{(z-1)r}{P_{S}\gamma_{sr}}}}{1+\frac{\gamma_{se}zr}{\gamma_{sr}}}
\ \nonumber\\
&+\binom {K-1}{K-2}\sum\limits_{r=0}^{K-1}\binom {K-1}{r} (-1)^{r}\frac{e^{-\frac{(z-1)(r+1)}{P_{S}\gamma_{sr}}}}{1+\frac{\gamma_{se}z(r+1)}{\gamma_{sr}}},
\end{align}
\end{small}
\begin{small}
\begin{align}
F_{Z_{t_2}}(z)=\left(1-\frac{e^{-\frac{z-1}{P_{R}\gamma_{rd}}}}{1+\frac{\gamma_{re}z}{\gamma_{rd}}}\right)^{K-1}\left(1+\binom {K-1}{K-2} \frac{e^{-\frac{z-1}{P_{R}\gamma_{rd}}}}{1+\frac{\gamma_{re}z}{\gamma_{rd}}}\right).
\end{align}
\end{small}
Since the computation of ${\mathbb E}[C_{SR,2}]$ is similar to that of ${\mathbb E}[C_{SR,1}]$, the computation of ${\mathbb E}[C_{RD,2}]$ is similar to that of ${\mathbb E}[C_{RD,1}]$, the details of the computation are omitted here. ${\mathbb E}[C_{SR,2}]$ is given by
\begin{small}
\begin{align}
&{\mathbb E}[C_{SR,2}]
\ \nonumber\\
&=\sum\limits_{r=1}^{K-1}\binom {K-1}{r}(-1)^{r}\frac{e^{\frac{r}{P_{S}\gamma_{sr}}}}{\ln2}
\ \nonumber\\
&\left[-E_{1}\left(\frac{r}{P_{S}\gamma_{sr}}\right)+e^{\frac{1}{P_{S}\gamma_{se}}}E_{1}\left(\frac{r}{P_{S}\gamma_{sr}}+\frac{1}{P_{S}\gamma_{se}}\right)\right]
\ \nonumber\\
&+\binom {K-1}{K-2}\sum\limits_{r=0}^{K-1}\binom {K-1}{r}(-1)^{r}\frac{e^{\frac{r+1}{P_{S}\gamma_{sr}}}}{\ln2}
\ \nonumber\\
&\left[-E_{1}\left(\frac{r+1}{P_{S}\gamma_{sr}}\right)+e^{\frac{1}{P_{S}\gamma_{se}}}E_{1}\left(\frac{r+1}{P_{S}\gamma_{sr}}+\frac{1}{P_{S}\gamma_{se}}\right)\right].\label{ECSR2}
\end{align}
\end{small}

And ${\mathbb E}[C_{RD,2}]$ can be obtained as
\begin{small}
\begin{align}
&{\mathbb E}[C_{RD,2}]
\ \nonumber\\
&=\sum\limits_{r=1}^{K-1}\binom {K-1}{r}(-1)^{r}\frac{\gamma_{re}e^{\frac{r}{P_{R}}\left(\frac{1}{\gamma_{rd}}+\frac{1}{\gamma_{re}}\right)}}{\ln2\gamma_{rd}}
\ \nonumber\\
&\Bigg[-e^{-\frac{r}{P_{R}\gamma_{re}}}E_{1}\left(\frac{r}{P_{R}\gamma_{rd}}\right)
\ \nonumber\\
&+\sum\limits_{i=1}^{r}\bigg(\frac{(-1)^{i+1}\left(\frac{r}{P_{R}\gamma_{re}}\right)^{i+1}E_{1}\left(\frac{r}{P_{R}}\left(\frac{1}{\gamma_{rd}}+\frac{1}{\gamma_{re}}\right)\right)}{(i-1)!}
\ \nonumber\\
&+\frac{e^{-\frac{r}{P_{R}}\left(\frac{1}{\gamma_{rd}}+\frac{1}{\gamma_{re}}\right)}}{\left(\frac{\gamma_{re}}{\gamma_{rd}}+1\right)^{i-1}}
\sum\limits_{j=0}^{i-2}\frac{(-1)^{j}\left(\frac{r}{P_{R}\gamma_{re}}\right)^{j}\left(\frac{\gamma_{re}}{\gamma_{rd}}+1\right)^{j}}{(i-1)(i-2)...(i-1-j)}\bigg)\Bigg]
\notag
\end{align}
\end{small}
\begin{small}
\begin{align}
&+\binom {K-1}{K-2}\sum\limits_{r=0}^{K-1}\binom {K-1}{r}(-1)^{r}\frac{\gamma_{re}e^{\frac{r+1}{P_{R}}\left(\frac{1}{\gamma_{rd}}+\frac{1}{\gamma_{re}}\right)}}{\ln2\gamma_{rd}}
\ \nonumber\\
&\Bigg[-e^{-\frac{r+1}{P_{R}\gamma_{re}}}E_{1}\left(\frac{r+1}{P_{R}\gamma_{rd}}\right)
\ \nonumber\\
&+\sum\limits_{i=1}^{r+1}\bigg(\frac{(-1)^{i+1}\left(\frac{r+1}{P_{R}\gamma_{re}}\right)^{i+1}E_{1}\left(\frac{r+1}{P_{R}}\left(\frac{1}{\gamma_{rd}}+\frac{1}{\gamma_{re}}\right)\right)}{(i-1)!}
\ \nonumber\\
&+\frac{e^{-\frac{r+1}{P_{R}}\left(\frac{1}{\gamma_{rd}}+\frac{1}{\gamma_{re}}\right)}}{\left(\frac{\gamma_{re}}{\gamma_{rd}}+1\right)^{i-1}}
\sum\limits_{j=0}^{i-2}\frac{(-1)^{j}\left(\frac{r+1}{P_{R}\gamma_{re}}\right)^{j}\left(\frac{\gamma_{re}}{\gamma_{rd}}+1\right)^{j}}{(i-1)(i-2)...(i-1-j)}\bigg)\Bigg].\label{ECRD2}
\end{align}
\end{small}

\quad $Computation$ $of$ $p_{12}$:
Given that $z_{r_1}>z_{r_2}$, $z_{t_1}>z_{t_2}$, we have
\begin{align}
p_{12}
&=P\{Q>0\}=P\{\min(z_{r_1},z_{t_2})>\min(z_{r_2},z_{t_1})\}
\ \nonumber\\
&=P\{z_{t_2}>z_{r_2}\}=\iint\limits_{z_{t_2}>z_{r_2}}f_{z_{r_2},z_{t_2}}(z_{r_2},z_{t_2})dz_{r_2}dz_{t_2}
\ \nonumber\\
&=\int_{0}^{\infty}f_{z_{r_2}}(z)(1-F_{z_{t_2}}(z))dz. \label{p12}
\end{align}
Since we have obtained the CDF of $z_{r_2}$ and $z_{t_2}$, we can express the probability of $Q>0$ , i.e., $p_{12}$, as an integral form and calculate its value numerically.

Note that ${\mathbb E}[C_{SR}]$ can be obtained by substituting (\ref{prop:ECSR1}), (\ref{ECSR2}) and $p_{12}$ into (\ref{ECSR}), ${\mathbb E}[C_{RD}]$ can be obtained by substituting (\ref{prop:ECRD1}), (\ref{ECRD2}) and $p_{12}$ into (\ref{ECRD}). Finally, the approximate closed-form expression of the secrecy throughput of the proposed scheme is obtained by substituting (\ref{ECSR}) and (\ref{ECRD}) into (\ref{secrecy throughput}).

Given the total power constraint SNR of the network, we can allocate the total power to the source and relays to achieve the best performance.

Regarding IFD-MRRS scheme, we need to allocate transmit energy to source and $K$ relays. The sources works for all time slots, therefore, we should have $(P_{S}+KP_{R}) \leq {\rm SNR}$. For max-link-ratio scheme, we should allocate transmit power to the source and $K$ relays for each time slot to enable each link to be capable of being selected for reception or transmission, so we should have $(P_{S}+KP_{R}) \leq {\rm SNR}$ as well. With max-min-ratio scheme, similarly, we should allocate transmit energy to the source and $K$ relays, albeit the data transmission occupies two time slots, so we should have $\frac{1}{2}(P_{S}+KP_{R}) \leq {\rm SNR}$.

Consider the derived expressions of secrecy throughput, once given the total power SNR, it is obvious that when $P_{S}$ is small, the throughput is limited by first hop. On the other hand, when $P_R$ is small, the second hop will be the bottleneck of the system. Therefore, there is always an optimal power allocation that maximizes the secrecy throughput.
\begin{Def}
The maximum secrecy throughput of IFD-MRRS is given by
\begin{align}
C_{max}=\max\limits_{(P_{S}+KP_{R}) \leq SNR}C_{IFD-MRRS}(P_{S},P_{R}).
\end{align}
\end{Def}
Similarly, we can define the maximum secrecy throughput for max-min-ratio scheme and max-link-ratio scheme.

\section{Numerical Results}
In this section, simulation results are given to verify the secrecy throughput for the proposed IFD-MRRS scheme. We assume that $\gamma_{sr} = \gamma_{rd} = \gamma_{se} = \gamma_{re} = 2$, unless specified otherwise.

\begin{figure}
    \centering
    \includegraphics[width=\figsize\textwidth]{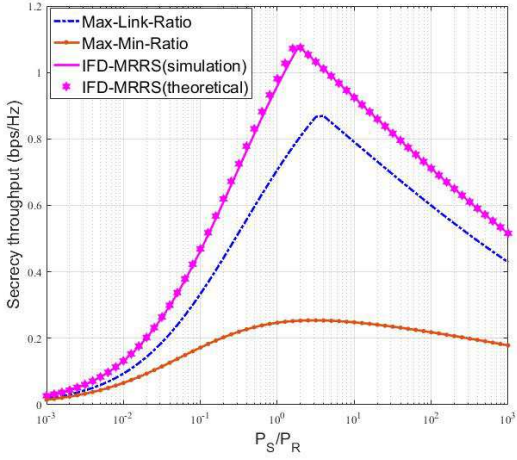}
    \caption{The secrecy throughput versus $P_{S}/P_{R}$ for different relaying protocols.}
    \label{figure2}
\end{figure}

Fig. \ref{figure2} plots the secrecy throughput versus $P_{S}/P_{R}$ for each scheme, where the relay number is set as $K=3$. We assume ${\rm SNR}=10$ dB. We can find that the secrecy throughput always has a peak value as $P_{S}/P_{R}$ varies, and the proposed scheme achieves the largest throughput. It is interesting that, each scheme does not achieve the maximum secrecy throughput when $\frac{P_{S}}{P_{R}}=1$, i.e., $P_{S}=P_{R}$. This is because, the distribution of the channels for the two hops is not symmetric, which can also be seen from the selection strategy (\ref{not symmetric}). We also note that the analytical results obtained based on the derivation in Section III are very close to the simulation results, which verifies the approximate closed-form expressions.

\begin{figure}
    \centering
    \includegraphics[width=\figsize\textwidth]{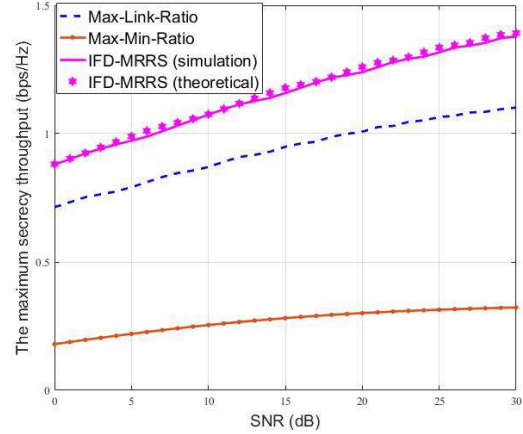}
    \caption{The maximum secrecy throughput versus SNR for different relaying protocols.}
    \label{figure3}
\end{figure}

In Fig. \ref{figure3}, we compare the maximum secrecy throughput of the proposed scheme with that of two existing max-ratio schemes as SNR varies, where the relay number is set as $K=3$. We can find that the proposed scheme achieves the best performance with power allocation. Also, we can find that the approximate expression holds for a wide range of SNR values.

\begin{figure}
    \centering
    \includegraphics[width=\figsize\textwidth]{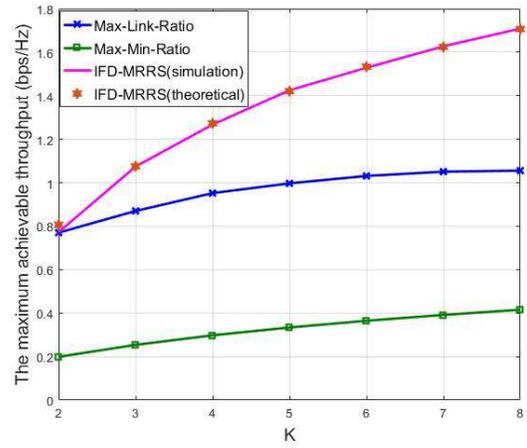}
    \caption{The maximum secrecy throughput versus different numbers of relays.}
    \label{figure4}
\end{figure}

Fig. \ref{figure4} plots the maximum secrecy throughput of each scheme versus the number of relays for ${\rm SNR}=10$ dB. We can find that the proposed scheme achieves the best performance in all cases. It is clearly shown that, the increase of the relay number can significantly improve the secrecy throughput performance.

\section{Conclusion}
In this paper, we have proposed an IFD-MRRS protocol for secure communications over buffer-aided cooperative relay networks. Notice that we considers the RF strategy such that the eavesdropper can only independently decode the signals received in the two hops. With the help of buffers at the relays, different relays for reception and transmission have been selected with the largest or the second largest ratio among $S-R$ and $R-D$ links. Approximate closed-form expressions for the secrecy throughput were derived. Numerical results in consistence with the analytical expressions show that the distribution of the channels of the two hop is not symmetric, and the proposed IFD-MRRS scheme achieves significantly higher secrecy throughput compared with two existing relay selection protocols for secure buffer-aided cooperative wireless networks.

\appendix
\subsection{Proof of expression (\ref{prop:ECSR1})}\label{app:ECSR1}
We first compute the CDF of $x$, the CDF of $x$ can be obtained as
\begin{align}
F_{X}(x)
&=P\left\{\max\{x_{k}\}\leq x\right\}
\ \nonumber\\
&=P\{x_{1}\leq x\}P\{x_{2}\leq x\}...P\{x_{K}\leq x\}
\ \nonumber\\
&=\left(1-e^{-\frac{x}{\gamma_{sr}}}\right)^{K}.
\end{align}
Since $x$ is independent of $y$, we have $f_{XY}(x,y)=f_{X}(x)f_{Y}(y)$.
Then the CDF of $z_{r_1}$ can be calculated as
\begin{small}
\begin{align}
&F_{Z_{r_1}}(z)
\ \nonumber\\
&=P\left\{\frac{1+P_{S}X}{1+P_{S}Y}\leq z\right\}=\iint\limits_{\frac{1+P_{S}x}{1+P_{S}y}\leq z}f_{XY}(x,y)dxdy
\ \nonumber\\
&=\int_{0}^{\infty}f_{Y}(y)\cdot F_{X}(x)|_{0}^{\frac{z}{P_{S}}+yz-\frac{1}{P_{S}}}dy
\ \nonumber\\
&=\int_{0}^{\infty}\frac{1}{\gamma_{se}}e^{-\frac{y}{\gamma_{se}}}\cdot \sum\limits_{r=0}^{K}\binom K r \left(-e^{-\frac{z-1}{P_{S}\gamma_{sr}}}e^{-\frac{yz}{\gamma_{sr}}}\right)^{r}dy
\ \nonumber\\
&=\sum\limits_{r=0}^{K}\binom K r (-1)^{r}\cdot\frac{e^{-\frac{(z-1)r}{P_{S}\gamma_{sr}}}}{\frac{zr\gamma_{se}}{\gamma_{sr}}+1}.
\end{align}
\end{small}
Then, we have
\begin{small}
\begin{align}
&{\mathbb E}[C_{SR,1}]
\ \nonumber\\
&={\rm log_{2}}z\left(F_{Z_{r_1}}(z)-1\right)\big|_{1}^{\infty}-\int_{1}^{\infty}\left(F_{Z_{r_1}}(z)-1\right)d({\rm log_{2}}z)
\notag
\end{align}
\end{small}
\begin{small}
\begin{align}
&=-\sum\limits_{r=1}^{K}\binom K r (-1)^{r}\frac{e^{\frac{r}{P_{S}\gamma_{sr}}}}{\ln2}\int_{1}^{\infty}
\left(\frac{e^{-\frac{zr}{P_{S}\gamma_{sr}}}}{z}-\frac{e^{-\frac{zr}{P_{S}\gamma_{sr}}}}{z+\frac{\gamma_{sr}}{\gamma_{se}r}}\right)dz
\ \nonumber\\
&=\sum\limits_{r=1}^{K}\binom K r (-1)^{r}\frac{e^{\frac{r}{P_{S}\gamma_{sr}}}}{\ln2}\bigg[-E_{1}\left(\frac{r}{P_{S}\gamma_{sr}}\right)
\ \nonumber\\
&+e^{\frac{1}{P_{S}\gamma_{se}}}E_{1}\left(\frac{r}{P_{S}\gamma_{sr}}+\frac{1}{P_{S}\gamma_{se}}\right)\bigg]\label{ECSR1case1}
\end{align}
\end{small}
where $E_{1}(x)=\int_{x}^\infty(e^{-t}/t)dt, x>0$ is the exponential integral function.

\subsection{Proof of expression (\ref{prop:ECRD1})}\label{app:ECRD1}
The CDF of $z_{k}$ can be obtained as
\begin{small}
\begin{align}
F_{Z_{k}}(z_{k})
&=P\left\{\frac{1+P_{R}X_{k}}{1+P_{R}Y_{k}}\leq z_{k}\right\}
\ \nonumber\\
&=\iint\limits_{\frac{1+P_{R}x_{k}}{1+P_{R}y_{k}}\leq z_{k}}f_{X_{k}Y_{k}}(x_{k},y_{k})dx_{k}dy_{k}
\ \nonumber\\
&=\int_{0}^{\infty}\int_{0}^{\frac{z_{k}}{P_{R}}+y_{k}z_{k}-\frac{1}{P_{R}}}f_{X_{k}Y_{k}}(x_{k},y_{k})dx_{k}dy_{k}
\ \nonumber\\
&=1-\frac{e^{-\frac{z_{k}-1}{\gamma_{rd}P_{R}}}}{1+\frac{z_{k}\gamma_{re}}{\gamma_{rd}}}.
\end{align}
\end{small}
Then the CDF of $z_{t_1}$ can be calculated as
\begin{align}
F_{Z_{t_1}}(z)
&=P\{Z_{t_1}\leq z\}=P\{\max\{Z_{k}\}\leq z\}
\ \nonumber\\
&=P\{Z_{1}\leq z\}\{Z_{2}\leq z\}...\{Z_{K}\leq z\}
\ \nonumber\\
&=\left(1-\frac{e^{-\frac{z-1}{P_{R}\gamma_{rd}}}}{1+\frac{z\gamma_{re}}{\gamma_{rd}}}\right)^{K}.\ \nonumber\\
\end{align}
After computation of $F_{Z_{t_1}}(z)$ and after some simplifications, ${\mathbb E}[C_{RD,1}]$ can be obtained as
\begin{small}
\begin{align}
&{\mathbb E}[C_{RD,1}]
\ \nonumber\\
&=-\sum\limits_{r=1}^{K}\binom K r (-1)^{r}\frac{e^{\frac{r}{P_{R}\gamma_{rd}}}}{\ln2}\int_{1}^{\infty}
\frac{e^{-\frac{zr}{P_{R}\gamma_{rd}}}}{z\left(1+\frac{\gamma_{re}}{\gamma_{rd}}z\right)^{r}}dz
\ \nonumber\\
&\xlongequal{u=1+\frac{\gamma_{re}}{\gamma_{rd}}z}-\sum\limits_{r=1}^{K}\binom K r (-1)^{r}\frac{e^{\frac{r}{P_{R}\gamma_{rd}}}}{\ln2}
\int_{\frac{\gamma_{re}}{\gamma_{rd}}+1}^{\infty}\frac{e^{-\frac{r(u-1)}{P_{R}\gamma_{re}}}}{\frac{\gamma_{rd}}{\gamma_{re}}(u-1)u^{r}}du
\ \nonumber\\
&=-\sum\limits_{r=1}^{K}\binom K r (-1)^{r}\frac{e^{\left(\frac{r}{P_{R}\gamma_{rd}}+\frac{r}{P_{R}\gamma_{re}}\right)}}{\ln2\frac{\gamma_{rd}}{\gamma_{re}}}
\ \nonumber\\
&\int_{\frac{\gamma_{re}}{\gamma_{rd}}+1}^{\infty}\left(\frac{e^{-\frac{r}{P_{R}\gamma_{re}}u}}{u-1}-\sum\limits_{i=1}^{r}\frac{e^{-\frac{r}{P_{R}\gamma_{re}}u}}{u^{i}}\right)du
\ \nonumber\\
&=\sum\limits_{r=1}^{K}\binom K r (-1)^{r}\frac{\gamma_{re}e^{\frac{r}{P_{R}}\left(\frac{1}{\gamma_{rd}}+\frac{1}{\gamma_{re}}\right)}}{\ln2\gamma_{rd}}\Bigg[-e^{-\frac{r}{P_{R}\gamma_{re}}}E_{1}\left(\frac{r}{P_{R}\gamma_{rd}}\right)
\ \nonumber\\
&+\sum\limits_{i=1}^{r}\bigg(\frac{(-1)^{i+1}\left(\frac{r}{P_{R}\gamma_{re}}\right)^{i+1}E_{1}\left(\frac{r}{P_{R}}\left(\frac{1}{\gamma_{rd}}+\frac{1}{\gamma_{re}}\right)\right)}{(i-1)!}
\ \nonumber\\
&+\frac{e^{-\frac{r}{P_{R}}\left(\frac{1}{\gamma_{rd}}+\frac{1}{\gamma_{re}}\right)}}{\left(\frac{\gamma_{re}}{\gamma_{rd}}+1\right)^{i-1}}
\sum\limits_{j=0}^{i-2}\frac{(-1)^{j}\left(\frac{r}{P_{R}\gamma_{re}}\right)^{j}\left(\frac{\gamma_{re}}{\gamma_{rd}}+1\right)^{j}}{(i-1)(i-2)...(i-1-j)}\bigg)\Bigg].\label{ECRD1case1}
\end{align}
\end{small}

\balance


\begin{thebibliography}{1}

\bibitem{5G1} M. Agiwal, A. Roy, and N. Saxena, ``Next generation 5g wireless networks: A comprehensive survey,'' \emph{IEEE Commun. Surveys Tuts.}, vol. 18, no. 3, pp. 1617-1655, Feb. 2016.

\bibitem{5G2} I. Ahmad, S. Shahabuddin, T. Kumar, J. Okwuibe, A. Gurtov, and M. Ylianttila, ``Security for 5G and beyond ,'' \emph{IEEE Commun. Surveys Tuts.}, vol. 21, no. 4, pp. 3682-3722, May 2019.

\bibitem{demand} T. Karygiannis and L. Owens, ``Wireless netork security,'' \emph{NIST Special Publication}, vol. 800, p. 48, Nov. 2002.

\bibitem{encryption} W. Stallings, \emph{Cryptography and Network Security: Principles and Practice}, 5th ed. Englewood Cliffs, NJ, USA: Prentice-Hall, Jan. 2010.

\bibitem{Wyner} A. D. Wyner, ''The wire-tap channel,'' \emph{Bell Syst. Tech. J.}, vol. 54, no. 8, pp. 1355-1387, 1975.

\bibitem{secrecy capacity} S. K. Leung-Yan-Cheong, M. E. Hellman, ``The Gaussian wiretap channel,'' \emph{IEEE Trans. Inf. Theory}, vol. 24, no. 4, pp. 451-456, Jul. 1978.

\bibitem{PHY} M. Bloch and J. Barros, \emph{Physical-Layer Security: From Information Theory to Security Engineering}. Cambridge, U. K.: Cambridge Univ. Press, 2011.

\bibitem{max-min-ratio} L. Lai and H. E. Gamal, ``The relay eavesdropper channel: Cooperation for secrecy,'' \emph{IEEE Trans. Inf. Theory}, vol. 54, no. 9, pp. 4005-4019, Sep. 2008.

\bibitem{AF/DF} L. Dong, Z. Han, A. P. Petropulu, and H. V. Poor, ``Improving wireless physical layer security via cooperating relays,'' \emph{IEEE Trans. Signal Process.}, vol. 58, no. 3, pp. 1875-1888, Mar. 2010.

\bibitem{RF} O. O. Koyluoglu, C. E. Koksal, and H. El Gamal, ``On secrecy capacity scaling in wireless networks,'' \emph{IEEE Trans. Inf. Theory}, vol. 58, no. 5, pp. 3000-3015, May 2012.

\bibitem{RF/DF} J. Mo, M. Tao, and Y. Liu, ``Realy placement for physical layer security: A secure connection perspective,'' \emph{IEEE Commun. Lett.}, vol. 16, no. 6, pp. 878-881, Jun. 2012.

\bibitem{Wan} J. Wan, D. Qiao, H. Wang, and H. Qian, ``Buffer-aided two-hop secure communications with power control and link selection,'' \emph{IEEE Trans. Wireless Commun.}, vol. 17, no. 11, pp. 7635-7647, Nov. 2018.

\bibitem{security-QoS} J. He, J. Liu, Y. Shen, X. Jiang, and N. Shiratori, ``Link selection for security-QoS tradeoffs in buffer-aided relaying networks,'' \emph{IEEE Trans. Inf. Forensics Secur.}, vol. 15, pp. 1347-1362, Sep. 2019.

\bibitem{buffer} X. Liao, Y. Zhang, Z. Wu, Y. Shen, X. Jiang, and H. Inamura, ``On security-delay trade-off in two-hop wireless networks with buffer-aided relay selection,'' \emph{IEEE Trans. Wireless Commun.}, vol. 17, no. 3, pp. 1893-1906, Mar. 2018.

\bibitem{adaptive} K. T. Phan, Y. Hong, and E. Viterbo, ``Adaptive resource allocation for secure two-hop relaying communication,'' \emph{IEEE Trans. Wireless Commun.}, vol. 17, no. 12, pp. 8457-8472, Dec. 2018.

\bibitem{cooperative} D. Wang, P. Ren, and J. Cheng, ``Cooperative secure communication in two-hop buffer-aided networks,'' \emph{IEEE Trans. Commun.}, vol. 66, no. 3, pp. 972-985, Mar. 2018.

\bibitem{max-link-ratio} G. Chen, Z. Tian, Y. Gong, Z. Chen, and J. A. Chambers, ``Max-Ratio relay selection in secure buffer-aided cooperative wireless networks,'' \emph{IEEE Trans. Inf. Forensics Secur.}, vol. 68, no. 4, pp. 719-729, Apr. 2014.

\bibitem{outage} X. Tang, Y. Cai, Y. Huang, T. Q. Duong, W. Yang, and W. Yang, ``Secrecy outage analysis of buffer-aided cooperative MIMO relaying systems,'' \emph{IEEE Trans. Veh. Technol.}, vol. 67, no. 3, pp. 2035-2048, Mar. 2018.

\bibitem{FD1} E. Sharma, R. Budhiraja, K. Vasudevan, and L. Hanzo, ``Full-Duplex massive MIMO multi-pair two-way AF relaying: energy efficiency optimization'' \emph{IEEE Trans. Commun.}, vol. 66, no. 8, pp. 3322-3340, Aug. 2018.

\bibitem{FD2} B. Chen, Y. Chen, Y. Chen, Y. Cao, Z. Ding, N. Zhao, and X. Wang, ``Secure primary transmission assisted by a secondary full-duplex NOMA relay'' \emph{IEEE Trans. Veh. Technol.}, vol. 68, no. 7, pp. 7214-7219, Jul. 2019.

\bibitem{mimick} A. Ikhlef, J. Kim, and R. Schober, ``Mimicking full-duplex relaying using half-duplex relays with buffers,'' \emph{IEEE Trans. Veh. Technol.}, vol. 61, no. 7, pp. 3025-3037, Sep. 2012.

\bibitem{average throughput} B. Xia, Y. Fan, J. Thompson, H. Vincent Poor, ``Buffering in a three-node relay network,'' \emph{IEEE Trans. Wireless Commun.}, vol. 7, no. 11, pp. 4492-4496, Nov. 2008.

\bibitem{order statistics} H. A. David and H. N. Nagaraja, \emph{Order Statistics}, 3rd ed. New York: Wiley, 2003.
\end{thebibliography}
\end{document}